\documentstyle[prd,aps]{revtex}
\begin{document}
\draft
\twocolumn[\hsize\textwidth\columnwidth\hsize\csname
@twocolumnfalse\endcsname
 \title{  Comments on  ``Analytic and Numerical Study of Preheating Dynamics''
}
\vskip 1cm
\author{Lev Kofman}
\address{Institute for Astronomy, University of Hawaii,
2680 Woodlawn Dr., Honolulu, HI 96822, USA}
\vskip .2cm
\author{ Andrei Linde}
\address{Department of Physics, Stanford University, Stanford, CA
94305, USA}
\vskip .2cm
\vskip .3cm
\author{ Alexei A. Starobinsky}
\address{Landau Institute for Theoretical Physics,
Kosygina St. 2, Moscow 117334, Russia}
\date{\today}
\maketitle
\begin{abstract}
In a recent paper  Boyanovsky et al (hep-ph/9608205)
 studied preheating for a particular
theory of the inflaton field.
 Although they results essentially confirm those of the
original papers on reheating after inflation,
 they  claimed that most of the earlier works on
preheating  did not take into account backreaction of created particles,
misused the Mathieu equation,  and were inconclusive about the symmetry
restoration after preheating.
 We explain why we cannot agree with these
statements.
\end{abstract}
\pacs{PACS: 98.80.Cq, 04.62.+v, 05.70.Fh \hskip 2cm SU-ITP-96-39, UHIFA96-45,
{}~hep-ph/9608341}
 \vskip2pc]

Recently there was an important progress in the theory of reheating of the
universe after inflation. As it was shown  in  \cite{KLS,Shtanov,Boyan1,Yosh},
   in many versions of
chaotic inflation   the  inflaton field $\phi$ very rapidly decays into its own
quanta and
other bosons due to  broad parametric
resonance.     To distinguish this stage of explosive reheating from the
stage of  particle decay and
thermalization, we called it {\it preheating}.
Investigation of this process should be
 performed   both in the broad and
in the narrow resonance regime, with an account taken of backreaction of
created particles and of expansion of the universe \cite{KLS}.
Bosons produced at that stage are far away from thermal equilibrium
and have enormously large occupation numbers.   Preheating leads to
many interesting effects. For example, specific nonthermal phase transitions
may occur soon after preheating, which are capable of restoring symmetry even
in the theories with symmetry breaking on the scale $\sim 10^{16}$ GeV
\cite{KLSSR,tkachev}.  Strong deviation from thermal equilibrium and the
possibility of production of superheavy particles  by oscillations of a
relatively light inflaton field may resurrect the theory of GUT baryogenesis
\cite{Kolb} and may considerably change the way baryons are produced in the
Affleck-Dine scenario \cite{Riotto}.

The theory of  preheating   is rather complicated and strongly model-depended.
Certain aspects of this theory were investigated by many authors, and a lot of
interesting information has been accumulated \cite{KLS}-\cite{Campbell}.
 However,
recently Boyanovsky,   de Vega,   Holman,  and   Salgado have written a paper
criticizing almost all other authors studying preheating \cite{BoyanNew}.
They write, in particular:
``There have been a number of papers (see refs.\cite{KLS,KLSSR,tkachev,Yosh})
dedicated to the analysis of the preheating process... In most treatments this
approximation neglects all back reaction
effects. We will argue that these approaches do {\bf not} agree with
our analytical and
numerical results and miss important physics... In fact, whereas the Mathieu
equation has
{\em infinitely many} forbidden bands, the exact equation has only {\em one}
forbidden band. Such an approximation, based on the chart of the unstable bands
of the Mathieu equation has been used in most \cite{KLS,Yosh}, etc.
treatments of preheating. These results must therefore be regarded as {\em
highly} suspect... We also discuss why the phenomenon of
symmetry restoration at preheating, discussed by various
authors\cite{KLSSR}-\cite{kolbriotto} is {\bf not} seen
to occur.''

We believe that these statements, made in  \cite{BoyanNew} in bold face and in
italic, are incorrect. They contradict the contents and misinterpret the
 papers they cited.

\

1. {\it Back reaction effects}\\
 Creation of particles leads to  several effects which can change  the
dynamics of the system.
Already in our first paper \cite{KLS} we did take into account backreaction of
created particles, which modifies  equations of motion for the interacting
fields.  We also took into account the contribution of the created particles to
the energy density of the universe, which controls the speed of the universe
expansion.
 Khlebnikov  and Tkachev  \cite{Khleb} studied the
feedback effects of
rescattering of created particles.
 A section of the review article by Kofman in \cite{KLS}
  is devoted to the  discussion of all of these
 feedback effects of created particles.
Ironically, only part of the feedback effects (modification of equation of
motion without an account taken of  expansion of the universe and
 rescattering) was considered in
\cite{BoyanNew}.

\

2. {\it The Mathieu equation}\\
  The main subject of investigation of ref.
\cite{KLS} was the theory of a massive inflaton field $\phi$  interacting with
 another scalar field $\chi$. Parametric resonance in this theory is
described explicitly  by the Mathieu equation,
 and comments of ref. \cite{BoyanNew} simply
do  not apply to this theory.

On the other hand, the main subject of investigation of   \cite{BoyanNew} was
the theory $m^2\phi^2/2 +\lambda\phi^4/4$.  There are two main regimes there:
$\phi \ll m/\sqrt \lambda$ and  $\phi \gg m/\sqrt \lambda$. The authors of
\cite{BoyanNew} do not make any distinction between these two regimes because
they neglect expansion of the universe. However, from  \cite{KLS} it
follows that  in expanding universe there is no parametric resonance at all for
$\phi \ll
m/\sqrt \lambda$. Therefore all results of   \cite{BoyanNew} related to
parametric resonance in this
regime do not give a correct description of reheating in the theory
$m^2\phi^2/2 +\lambda\phi^4/4$.

As for the regime $\phi \gg m/\sqrt \lambda$, the main statement of ref.
\cite{BoyanNew} is that the parametric resonance in this model is described by
  Lame equation rather than by Mathieu equation.
 However, this fact was already emphasized earlier
in   \cite{KLS}. To study parametric resonance in this theory,  one can
approximate
  Lame equation  by Mathieu equation.
 The resonance which was  found in \cite{KLS}
for this particular theory
was only in one band, so the statement of ref. \cite{BoyanNew} that we used
charts of many bands is   misplaced.

Approximate treatment of Lame equations in terms of Mathieu equation was made
in \cite{KLS} only as a first illustrative part of a complete investigation.
It was found \cite{KLS} that fluctuations approximated by
Mathieu equation the unstable modes grow as
$e^{3.4\mu \sqrt\lambda \phi t}$, where $\mu \sim 0.07$.
Direct investigation of the Lame equation
reveal similar
result but with $\mu \sim  0.04$, see the review  by Kofman in \cite{KLS}. This
was the value which was used in our subsequent paper  \cite{KLSSR}.
 Khlebnikov  and Tkachev  \cite{Khleb}
made a more complete numerical investigation of preheating
 in the theory $\lambda\phi^4/4$ than the investigation performed in
\cite{BoyanNew}, because they took into account rescattering of produced
particles and expansion of the universe. They obtained the same value of $\mu$
\cite{Khleb}.
All of these results were ignored by the authors of ref. \cite{BoyanNew}.

A more general theory studied in \cite{KLS}   contains both the terms
$m^2\phi^2/2 +\lambda\phi^4/4$ and $g^2\phi^2\chi^2$. In the limit $\phi \ll
m/\sqrt \lambda$
 decay due to parametric resonance can go  only to the field $\chi$, and there
are many
instability bands in this regime. As for the regime $\phi \gg m/\sqrt \lambda$,
$g^2 \gg  \lambda$,
the number of instability bands for the field $\chi$ is infinitely large for
all but very special relations between $g^2$ and $\lambda$.

It is instructive to consider the maximum value of
 growth index  $\mu$ in the unstable bands
 as a function of
$g^2/ \lambda$.
It turns out that the model considered in \cite{BoyanNew}
 in a certain sense
 is a degenerate case.
 The function  $\mu (g^2/\lambda)$ has a minimum
at $g^2/ \lambda=3$. In this case equation describing creation of
$\chi$-particles is similar to the equation for $\phi$-particles in the theory
$\lambda\phi^4/4$. In this particular case there is a single instability band.
However,
in the regime $g^2 \gg \lambda$  there are many instability bands, and the
results for $\mu$ are converging to
 those obtained from the  Mathieu equation.

\

3. {\it Symmetry restoration from preheating}\\
Finally,  let us consider the theory of nonthermal phase
transitions \cite{KLSSR,tkachev}. These phase transitions have been found in
several versions of chaotic
inflation theory when the inflaton field $\phi$ rolls down from very large
values $\phi \sim M_p$. Then the oscillations of this field produces
enormously large fluctuations $\langle\phi^2\rangle$ and
$\langle\chi^2\rangle$, which are capable of restoring symmetry in the theory.

The authors of ref. \cite{BoyanNew} claim that they did not find any
symmetry restoration from preheating.
 However, they  studied a  regime which is completely different from
that of \cite{KLSSR,tkachev}.
Instead of analysing fluctuations produced by the field rolling from $\phi \sim
M_p$, they studied the theory where the field $\phi$ falls to the minimum of
the effective potential from the vicinity of $\phi = 0$.  Symmetry restoration
in
this case would imply return of the field $\phi$ exactly to the point $\phi =
0$, which is hardly possible because it looses some part of its
energy for particle production. Thus it is
not surprising that Boyanovsky {\it et al}  did not
find in their model
 the effect of symmetry restoration discovered in \cite{KLSSR,tkachev} in a
different context.

\end{document}